\documentclass[aps, prl, twocolumn]{revtex4-1}
\pdfoutput=1

\usepackage{amsmath}
\usepackage{amssymb}
\usepackage{color}

\usepackage{hyperref}
\usepackage{verbatim}
\usepackage{graphicx}
\usepackage{subfig}

\newcommand{\bi}{\begin{itemize}}
\newcommand{\ei}{\end{itemize}}
\newcommand{\be}{\begin{equation}}
\newcommand{\ee}{\end{equation}}

\newcommand{\bml}{\begin{multline}}
\newcommand{\emll}{\end{multline}}

\newcommand{\nn}{\nonumber}

\def\({\left(} \def\){\right)}
\def\[{\left[} \def\]{\right]}

\def\al{\alpha}

\def\mO{\mathcal{O}}

\def\mA{\mathcal{A}}

\def\a{\alpha}

\def\lam{\lambda}

\newcommand{\G}{\Gamma}

\def\d{\partial}

\newcommand{\la}{\langle}
\newcommand{\ra}{\rangle}

\newcommand{\bea}{\begin{eqnarray}}
\newcommand{\eea}{\end{eqnarray}}

\def\ie{\begin{equation}\begin{aligned}}
\def\fe{\end{aligned}\end{equation}}

\def\tilde{\widetilde}
\def\t{\tilde}

\def\d{\partial}

\def\1{{\mathds 1}}

\def\mA{\mathcal{A}}

\usepackage[UKenglish]{datetime}

\begin{document}

\title{ Chaos in  a many-string scattering amplitude }

\author{Vladimir Rosenhaus \vspace{.1cm}}

\affiliation
{Initiative for the Theoretical Sciences, \\The Graduate Center, City University of New York \\
365 Fifth Ave, New York, NY 10016}

\begin{abstract}
String theory provides a compact integral expression for the tree-level scattering amplitude of an arbitrary number of light strings. We focus on amplitudes involving a few tachyons and many photons, with a special choice of polarizations and kinematics. We pick out a particular pole in the amplitude --  one corresponding to successive photon scatterings, which lead to an intermediate state with a highly excited string in a definite state. This provides a physical process  which creates a highly excited string. The observed erratic behavior of the amplitude suggests that this may serve as a simple and explicit illustration of chaos in many-particle scattering.

\end{abstract}
\maketitle

\section{Introduction}

\vspace{-.2cm}
The essence of classical few-body chaos is simple to state: under time evolution, a region of phase space undergoes repeated stretching and folding. A small and smooth patch of phase space evolves into a highly intricate and complex structure which spans a vast region of phase space, while retaining the same volume as the initial region. As a result, late time observables behave erratically as functions of the initial conditions. \\[-10pt]

A similarly simple and clear -- and striking -- picture of chaos in quantum field theory has been lacking~\cite{Note1}.  In  \cite{VRchaos} we proposed that one look for signatures of chaos in the erratic behavior of scattering amplitudes of a large number of particles. 
In this note we will show that this diagnostic is satisfied in weakly coupled bosonic string theory. We will look at amplitudes involving a few tachyons and a large number of photons, with particular kinematics. We will show that the amplitude is highly erratic under a small change in the momentum of one of the tachyons or photons. \\[-10pt]

An amplitude involving a large number of particles is in general difficult to analyze and, moreover, it is difficult to know for which kinematics  to expect chaos. Here we carefully pick the kinematics so that the intermediate state is a highly excited string. In \cite{GRhigh} we studied scattering amplitudes involving generic highly excited strings, showing that they exhibit erratic behavior; an intuitively plausible result. We claimed that this indicates chaos in a many-photon amplitude, by arguing that the DDF construction \cite{DDF, Skliros} of the excited string vertex operator has a physical interpretation of forming an highly excited string by repeatedly scattering photons off of an initial tachyon. Here we verify this claim, by obtaining the many-photon amplitude directly, for a special choice of photon polarizations.

\section{String scattering amplitudes}
\vspace{-.2cm}
String scattering amplitudes are given by integrals of correlation functions of vertex operators $V(z_i)$ over the locations $z_i$ on the string worldsheet \cite{Polchinski}, 
\be
\mA = \frac{1}{\text{vol}(SL_2)} \int d z_i\ \la \prod_i V(z_i)\ra~,
\ee
where the vertex operators are functions of the string location $X^{\mu}(z)$, where $\mu$ runs over the $D$ dimensional ambient spacetime. The string field satisfies the Polyakov action, 
\be
-\frac{1}{2\pi } \int d \tau d\sigma\, \sqrt{-\gamma} \gamma^{a b} \d_a X^{\mu} \d_b X_{\mu}~,
\ee
where we have set $\al' = 1/2$. We may fix the worldsheet metric $\gamma_{ab}$ to be  flat. The $X^{\mu}$ then become free fields, with the correlation function, 
\be \label{xxcc}
\la X^{\mu} (z_1) \d X^{\nu}(z_2)\ra = \frac{ \eta^{\mu\nu}}{z_{12}}~.
\ee

We will be interested in amplitudes involving tachyons and photons. The tachyon vertex operator is the familiar $:\!e^{i p \cdot X(w)}\!\!:$~,  where $p^{\mu}$ is the tachyon momentum, $w$  is the worldsheet coordinate,  and the colons denote normal ordering. The photon vertex operator is,
\be
:i\, \zeta {\cdot} \d X(z) e^{i p \cdot X(z)}:~,
\ee
where the polarization $\zeta$ is orthogonal to the momentum, $\zeta {\cdot} p = 0$. A trick for working with the photon vertex operator is to write it as \cite{GSW},
 \be
: e^{i \zeta \cdot \d X(z)}:\, : e^{i p \cdot X(z)}:~
 \ee
 and keep the linear in $\zeta$ term. 
 
 The correlation function is now straightforward to compute -  through Wick contraction and use  of (\ref{xxcc}) -  and
 the amplitude involving any number of tachyons and  any number of photons immediately follows,
 \bml \label{AmpTP}
\mA =\frac{1}{\text{vol}(SL_2)} \int  d w_i  d z_a\,  \prod_{i<j} w_{i j}^{p_i \cdot p_j}\prod _{a<b} z_{ab}^{p_a \cdot p_b}\\
 \prod_{i,a} (w_i - z_a)^{p_i \cdot p_a}
 \exp\(\sum_{a \neq b}\frac{1}{2} \frac{\zeta_a {\cdot} \zeta_b}{z_{ab}^2} - \sum_{a, i}\frac{p_i {\cdot} \zeta_a}{w_i {-} z_a}\)\Big|_{\mO(\zeta_a)}
\end{multline}
where $z_{i j} \equiv z_i {-} z_j$,   the indices $i,j$ runs over the tachyons, the indices $a, b$ runsover the photons, and we are keeping  the linear in $\zeta_a$ term for each $a$. 
\\

\noindent \textit{Veneziano Amplitude: }The most familiar case is scattering of four tachyons and no photons - the Veneziano amplitude, 
\be
\mA = \frac{1}{\text{vol}(SL_2)}\int d w_i \prod_{1\leq i<j\leq 4}|w_{i j}|^{p_i \cdot p_j}~.
\ee
We may use SL$_2$ symmetry to fix the location of three of the four points. However the symmetry cannot be used to change the ordering of the points, so we must sum over the orderings.  Fixing $w_1=0$, $w_2=1$, $w_3 = \infty$, we get, 
\be
\mA =2 \int_{-\infty}^{\infty} d w_4 |w_4|^{p_1\cdot p_4} |1-w_4|^{p_2\cdot p_4}~.
\ee
There were $6$ distinct orderings, but they are equal in pairs; hence the factor of $2$. 
Each of the three orderings: $-\infty{<}w_4{<}0$, $\, 0{<}w_4{<}1$, and $1{<}w_4{ <}\infty$ gives a beta function, with the result, 
\be \nn
\mA = 2\beta( p_1 \cdot p_4 +1, p_2 \cdot p_4+1)+ (p_2\leftrightarrow p_4)+ (p_3\leftrightarrow p_4)~.
\ee
The amplitude has poles in the $s$, $t$, and $u$ channels. In the $s$ channel there are poles whenever $p_1 \cdot p_2 = - (N+1)$, with positive integer $N$. This corresponds to the exchange of an excited string of mass $m^2 = 2(N-1)$. 

\subsection{Three tachyons and $J$ photons}
\begin{figure}
\centering
\includegraphics[width=2.3in]{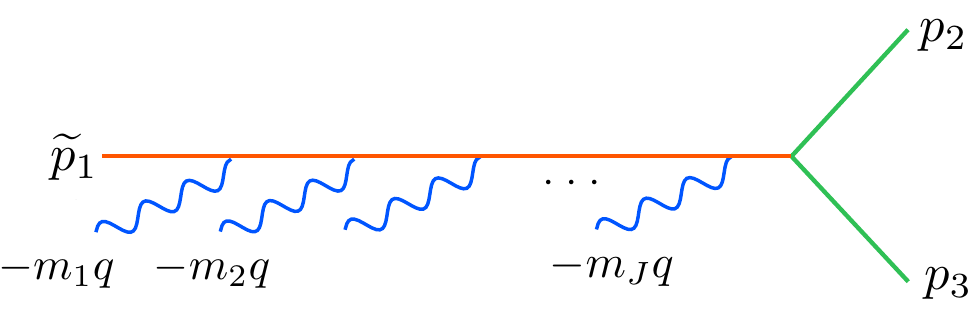}   
\caption{   
We look at an amplitude containing three tachyons and $J$ photons. We are interested in the piece of the amplitude in which the photons (blue) with  momenta $- m_a q$ scatter off of the initial tachyon (red) with momentum $\t p_1$, eventually decaying into two tachyons (green) with momenta $p_2$ and $p_3$. }
\label{excitedDecay}
\end{figure}
The amplitude (\ref{AmpTP}) with a general number of tachyons and photons, in a general kinematic configuration, is challenging to write in a more explicit form. We will consider a particular kinematic configuration,  involving three tachyons and $J$ photons, see Fig.~\ref{excitedDecay}. The tachyons have momenta, $\t p_1\equiv  p _1 + N q$, $p_2$, and $p_3$, where $q$ is null vector. The photons have  polarizations $\lam_a$ and momenta $- m_a q$ where $m_a$ is an integer and the index $a = 1, 2,\ldots J$. In addition, $N = \sum_{a=1}^J m_a$.  There are many different tree level processes that can occur in this amplitude. We are interested in the one in which the initial tachyon $\t p_1$ undergoes successive collisions with the photons, eventually decaying into two tachyons. Moreover, to make (\ref{AmpTP}) tractable we take the photon polarizations to be orthogonal to each other, $\lam_a \cdot\lam_b = 0$ for all $a,b$. A simple way of achieving this is to have all the $\lam_a$ be equal to each other, and equal to  transverse circular polarization, $(1,i)$ in the transverse direction. The amplitude (\ref{AmpTP}) then reduces to,
\bml \label{118}
\mA =\frac{-1}{\text{vol}(SL_2)} \int\! d w_i d z_a\,  w_{12}^{\t p_1 \cdot p_2 } w_{13}^{\t p_1 \cdot p_3} w_{23}^{p_2 \cdot p_3} \\\prod_{a=1}^J\(\sum_{i=1}^{3} \frac{p_i {\cdot} \lam_a}{w_i {-} z_{ a}}\)
 \frac{ (w_2 {-} z_a)^{- m_a p_2 \cdot q}}{(w_1 {-} z_a)^{m_a \t p_1 \cdot q} (w_3 {-}z_{a})^{m_a p_3 \cdot q}}
\end{multline}

The amplitude should have a pole when any of the internal string states are on shell. We look at the $J$'th order pole, when all the internal states are on shell. The intermediate momenta are given by $\t p_1 {-}\sum_{i=1}^a m_i q$, and are on shell if the square of the momentum is equal to the square of the mass of an excited string, $2(\sum_{i=1}^a m_i-1)$. The amplitude is of the form, 
\be \label{119}
\mA = \mA_{H\rightarrow T + T}\prod_{a=1}^J\! \frac{1}{(\t p_1 {-}\sum_{i=1}^a m_i q)^2 {+} 2(\sum_{i=1}^a m_i{-}1)} + \ldots
\ee
Simplifying, we may write this as, 
\be \label{eq10}
\mA =  \mA_{H\rightarrow T + T}\frac{1}{\prod_{a=1}^J\sum_{i=1}^a 2 m_i} \frac{1}{(1{-}\t p_1 \cdot q)^J}+\ldots
 \ee
In  Appendix~\ref{appA} we pick out the pole, finding that the residue is, 
\be \label{eq11}
\mA_{H\rightarrow T + T} \sim \prod_{a=1}^J p_3{ \cdot} \lam_a\, P_{m_a}(p_3{ \cdot} q)~,
\ee
where the function $P_m(x)$ is defined as, 
\be  \label{Pma}
P_m(x) = \frac{(1+m x)(2+ m x) \cdots (m{-}1 + m x)}{(m-1)!}~.
\ee

\section{Excited Strings}
As the notation suggests, the residue of the amplitude, $\mA_{H\rightarrow T + T}$, is in fact the amplitude for an excited string to decay into two tachyons. This was found in \cite{GRhigh}. The  excited string is in the state, 
\be
(\lam_1{\cdot} A_{-m_1}) (\lam_2{\cdot} A_{-m_2}) \cdots (\lam_J {\cdot} A_{- m_J}) |0;\t p_1\ra~,
\ee
where $A_{-k}$ is the DDF creation operator  \cite{DDF, Skliros} which excites the $k$'th mode of a string, and $|0; \t p_1\ra$ denotes an unexcited string (a tachyon) with center of mass momentum $\t p_1$. The mass of the excited string is $m^2 = 2(N-1)$ where $N$ is the level, $N = \sum_i m_i$, and the  $m_i$  appeared above in parameterizing the momenta of the photons. 

Let us discuss the reason for this equality. The most straightforward way to find an amplitude involving an excited string is through use of the corresponding vertex operator.  The vertex operator for an excited string of momentum $p$ is some polynomial of derivatives of $X$, $\d^{k} X^{\mu}$, multiplying $e^{i p\cdot X}$. The polynomial will have a total of $N$ derivatives, for a vertex operator of a string at level $N$. The polynomial is fixed by the requirement that the vertex operator  satisfy the Virasoro constraints;  however, solving these constraints is challenging. 

In more detail, the location of an open string is denoted by $X^{\mu}(\sigma, t)$, with the index $\mu$ running over the $D$ spacetime dimensions, and $\sigma$ parameterizing the coordinate along the string with endpoints $\sigma = 0,\, \pi$. For each $\mu$,  $X^{\mu}(\sigma, t)$ satisfies the wave equation,  $(- \d_t^2 + \d_{\sigma}^2)X^{\mu}(\sigma, t) = 0$, whose solution in terms of Fourier modes is, 
\be  \label{Xmu}
X^{\mu}(\sigma, t)  = x^{\mu} + p^{\mu} t + \sum_{n\neq 0} \frac{1}{n} \al_{n}^{\mu} e^{- i nt} \cos n \sigma~,
\ee
with Fourier coefficients $\a_n^{\mu}$ for integer $n$. The center of mass position and momentum are denoted by  $x^{\mu}$ and $p^{\mu}$, respectively, and the string length has been set to one.

Upon quantizing, the Fourier coefficients are promoted to operators, with commutation relations $\[\al_m^{\mu}, \al_n^{\nu} \] = m \delta_{m+n} \eta^{\mu \nu}$. For each mode $m>0$ and each  direction $\mu$, there is a harmonic oscillator of frequency $m$, with  creation operator $\al_{-m}^{\mu}$ and annihilation operator $\al_{m}^{\mu}$.  For a fundamental string in string theory, there is an additional caveat: the coordinates $\sigma$ and $t$ were chosen arbitrarily, yet one should be allowed to perform a change of coordinates, without affecting any physical answer. This  gives the constraint that the worldsheet energy-momentum tensor (not to be confused with the spacetime energy-momentum tensor) vanishes for physical states. This translates into relations  among  the $\al_{n}^{\mu}$ operators; these are the Virasoro constraints. 

The most efficient way to find the vertex operator for an excited string is not to solve the Virasoro constraints, but rather to construct the operator within the DDF formalism  \cite{DDF, Skliros}, see also \cite{Skliros0, GRhigh, Bianchi, DHoker}. Within the DDF construction, one starts with a tachyon vertex operator, takes the OPE with a photon vertex operator, picks out the pole, and then repeats the procedure with successive photon vertex operators.

\subsubsection{Many photons  versus an excited string}
The DDF construction mirrors the physical  process of starting with a tachyon and successively scattering photons off of it, and after each scattering event picking out the intermediate string state that is on shell. It is clear that if one wants the amplitude involving excited strings, it is generally more efficient to use the DDF operator, rather than picking out the relevant pole of a many tachyon/photon amplitude, as we did in order to obtain (\ref{eq10}). 

It would generally be difficult to study an amplitude with $3$ tachyons and $J$ photons, with large $J$, because in addition to the process we are interested in -- of the photons successively scattering off of the tachyon, the amplitude has pieces in which the photons interact with each other. 
The reason for the simplicity in the amplitude (\ref{eq10}) was that we took all the photon polarizations to be orthogonal to each other, thereby preventing this from occurring. More technically, had we not taken the photon polarizations to be orthogonal to each other, the amplitude with multiple photons (\ref{AmpTP}) would have had terms involving $\lam_a \cdot \lam_b/z_{ab}^2$. The integrals over the different photon insertion points would no longer decouple. For the general case, in which the photon polarizations are not orthogonal, it is better to compute the amplitude using the DDF operators, which is what we did in \cite{GRhigh}.

\section{Decay of a generic excited string}
Let us look in more detail at the amplitude $\mA_{H\rightarrow T + T}$ (\ref{eq11})
for the decay of an excited string into two tachyons \cite{GRhigh}. For simplicity, we take all the photons forming the excited string to have the same polarization $\lam$. The state is $\prod_{m=1}^{\infty} (\lam \cdot A_{-m})^{n_m} |0; \t p_1\ra$, where we can now parameterize the state by the excitation levels $\{n_m\}$ of mode $m$. The total level of the string is $N = \sum_{m=1}^{\infty} n_m m$. 

We  take concrete kinematics.  The initial tachyon has momentum  $\t p_1 = N q+ \sqrt{2N} \, (1,0,0,0)$. The photons have momenta proportional to $q$, where $q = -\frac{1}{\sqrt{2N}}(1,\sin \beta,  \cos \beta,0)$. The photon polarizations are $\lam = \frac{1}{\sqrt{2}}(0, - \cos \beta, \sin \beta,  i)$. After the photons scatter off of the initial tachyon, the resulting excited string decays into two tachyons, with momenta $
p_2 = - \frac{\sqrt{2N}}{2}(1,\sin \theta,  \cos \theta,0)$
and $p_3 = - \frac{\sqrt{2N}}{2}(1,- \sin \theta, - \cos \theta,0)$.

The amplitude  $\mA_{H\rightarrow T + T}(\al)$ is a function of the specific state of the excited string, parameterized by the excitation levels of the modes $\{n_m\}$, and of the difference in angle $\al = \theta - \beta$ between the photons used to create the excited string and the outgoing photons. The amplitude (\ref{eq11}) becomes \cite{GRhigh},  
\be \label{mA}
\mA_{H\rightarrow T + T}(\al) \sim \prod_{m=1}^{\infty} \( \sin \al\,  P_m(- \cos^2 \frac{\al}{2}) \)^{n_m}~, 
\ee
where $P_m(a)$ was given in (\ref{Pma}). 
For $-1<a<0$, the range relevant for us, $P_m(a)$  oscillates as a function of $a$. 

If only the low lying modes of the string are excited, $n_m \neq 0$ for small values of $m$, the amplitude is simple. As higher modes are excited, the amplitude develops more structure. We are interested in the  decay of a generic highly excited string.  
 For a generic state and for $N \gg 1$,  most of the modes  making up the mass of the excited string have a frequency of order $\sqrt{N}$ \cite{GRhigh}: a large number for a highly excited string. We may apply Stirling's approximation to $P_m(a)$ defined in (\ref{Pma}) for $m |a| \gg 1$. We are only interested in the oscillatory part, which behaves as $P_m(a) \sim - \sin (\pi m a) \cdots $. 
The oscillatory part of the amplitude is therefore of the form,
\be \label{mAal}
\mA_{H\rightarrow T + T}(\al)\sim ( \sin \al)^J\prod_{m=1}^{\infty} \sin\(\pi m \cos^{2}\frac{\al}{2}\)^{n_m}~,
\ee
where $J = \sum_{m=1}^{\infty} n_m$ is the spin of the string. The result (\ref{mAal}) is remarkably simple.

\section{Erratic behavior in string scattering}
\begin{figure}
\centering
\includegraphics[width=3in]{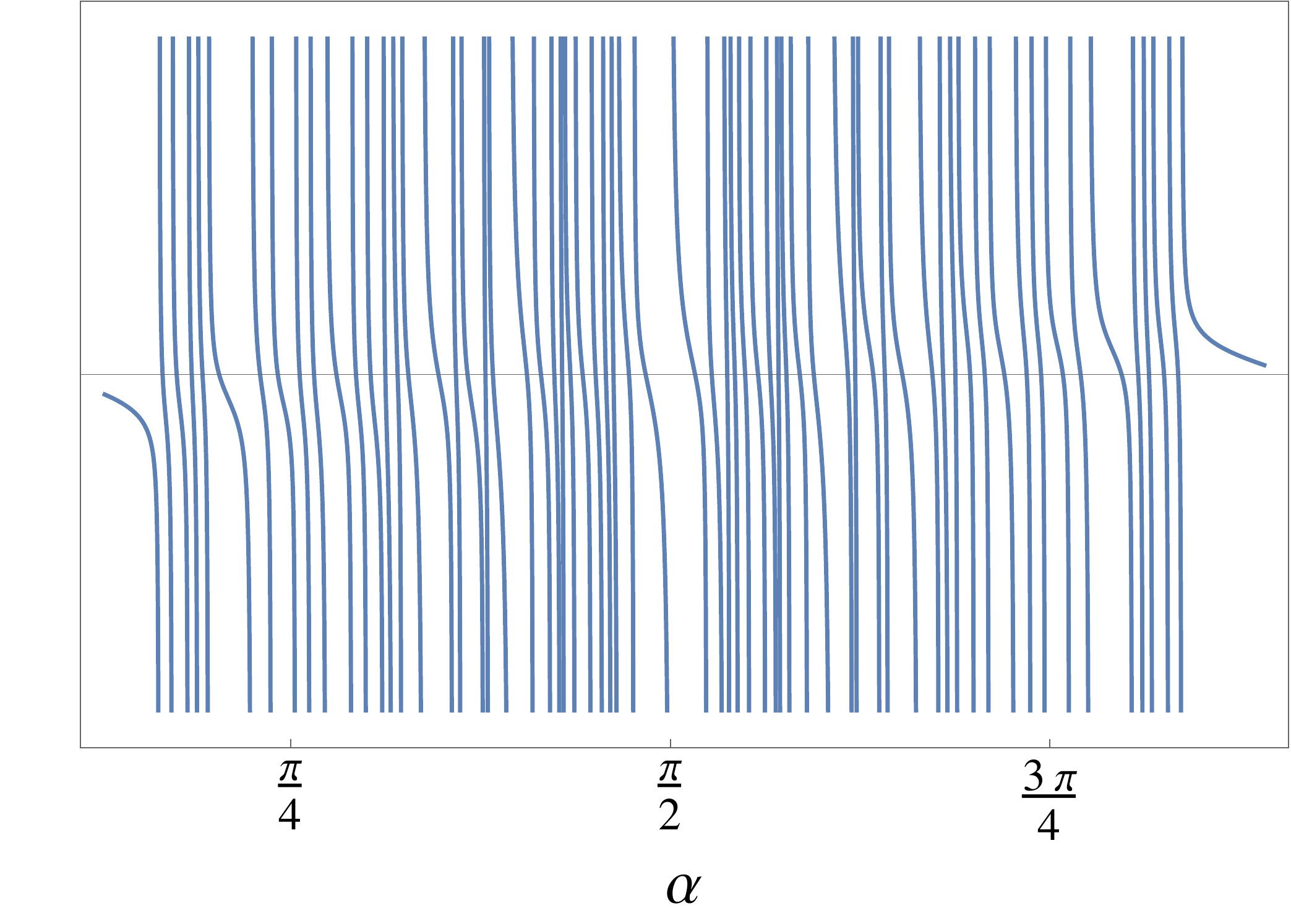}   
\caption{A plot of the derivative of the logarithm of the amplitude (\ref{mAal}), $d \log \mA_{H\rightarrow T + T}(\al)/d\al$, for a generic excited string state at level $N=100$. The partition of $N$ used in the plot is $(18, 16, 14, 13, 12, 9, 5, 4, 3, 3, 1, 1, 1)$.}\label{plot2}
\end{figure}
Is this sector of string theory chaotic? There is no generally accepted notion of what chaos in quantum field theory or string theory means. In \cite{VRchaos} we proposed that chaos be diagnosed  by erratic behavior of a many-particle amplitude, under a change in the momentum of one of the particles. 

Our  amplitude involves $J+3$ particles: an initial tachyon, $J$ photons sent in at relative angle $\beta$ (with $n_m$ photons of energy $m$), and two outgoing tachyons at angles $\theta$ and $-\theta$. A change in the momentum of one of the particles can either be a change in its direction or in its magnitude. We need to see if the amplitude is erratic under a small change in $\theta$ or $\beta$, or a small change in the occupation numbers $n_m$. 

We repeat the argument given in \cite{GRhigh}. The amplitude $\mA_{H\rightarrow T + T}(\al)$ (\ref{mAal}) is a function of $\al = \theta -\beta$ and has a zero at each angle $\al$ for which $\cos^2 \frac{\al}{2} = \frac{j}{m}$ for integer $j$ less than $m$, and for all $m$ that are excited (having a nonzero occupation number $n_m$). 
In the limit of infinite $N$, and nonzero $n_m$ for all $m$, the amplitude will have a zero  for every angle $\al$ at which $\cos^2 \frac{\al}{2}$ is a rational number. For finite $N$, only some of the modes are excited. Consider some generic excited string state, with $n_m$ found by considering a random partition of $N$ for large $N$. Let $m$ be some mode for which $n_m$ is nonzero and $n_{m+1}$ is zero. Now perform a small change in the state, setting $n_m$ to zero and $n_{m+1}$ to 1, while leaving the occupation numbers of all other modes unchanged. The zeros in the amplitude that were present at  $\cos^2 \frac{\al}{2} = \frac{j}{m}$ are gone, and there are now zeros at  $\cos^2 \frac{\al}{2} = \frac{j}{m+1}$. The amplitude has undergone a large change. 

Consider a small change in the angle $\al$. The change in the amplitude $\mA_{H\rightarrow T + T}(\al)$ is controlled by the derivative of the logarithm of $\mA_{H\rightarrow T + T}(\al)$.
A plot for some particular  generic state is shown in Fig.~\ref{plot2}. For a generic state, the number of zeros in the amplitude scales as $N$ for large $N$. The density of zeros per unit angle therefore scales as $1/N$. The distribution of zeros is controlled by the precise choice of $\{n_m\}$.
For some generic excited string, the distribution looks erratic. 


\vspace{-.15cm}

\section{Discussion}

For a general quantum field theory, it is likely that seeing chaos in scattering requires having a large number of closely spaced resonances, and correspondingly strong coupling. Weakly coupled string theory is an exception, with a free highly excited string already having an enormous number of internal states. Indeed, the same tree level scattering experiment studied here, when looked at in the context of, for instance, QED, would have been uninteresting: photons scattering off of an electron still give back an electron. The key here is that a photon scattering off of a string gives an excited string in a superposition of many different states. The chaos in string scattering may be related to the chaos in black hole scattering \cite{Shenker:2013pqa, Kitaev, PolchinskiC}  through the correspondence principle  between black holes and strings \cite{HorowitzPolchinski, Amati, Chen:2021dsw}. 

There are hopefully other tractable examples in which many-particle scattering amplitudes in  quantum field theory exhibit erratic behavior. One would  like to  be able to quantify how chaotic the $S$-matrix is. For semiclassical processes one can compute a Lyapunov exponent through the out-of-time-order correlator \cite{LO, Kitaev, MSS}, yet to have a generally valid diagnostic one should presumably only make use of the $S$-matrix. For chaotic scattering in quantum mechanics it has been proposed that the $S$-matrix has features of a random unitary matrix \cite{Blumel3, Smilansky}; for application to quantum field theory, this diagnostic would need to be refined. There have been extensive studies of matrix elements of operators in quantum many-body systems,  behavior embodied in the eigenstate thermalization hypothesis \cite{ETH, ETH2, Rigol, DAlessio:2015qtq,  ETH3, Kraus}. One may hope to likewise understand the behavior of  many-particle $S$-matrices, perhaps in part on the basis of fundamental properties of quantum field theories \cite{Paulos:2016but, Paulos:2017fhb, Homrich:2019cbt, Guerrieri:2021ivu, Correia:2021etg, Correia:2020xtr, Mizera:2019ose, Huang:2020nqy, Arkani-Hamed:2020blm}. 
\\

\begin{acknowledgments}

\noindent\textit{Acknowledgments:} I am grateful to D.~Gross for collaboration on closely related work.  I thank N.~Arkani-Hamed, S.~Mizera, and A.~Tseytlin for helpful discussions. This work was supported by the ITS. 
\end{acknowledgments}

\appendix

\section{Appendix A: Extracting the Pole } \label{appA}


Here we show that the amplitude (\ref{118}), near the $J$'th order pole, has the behavior (\ref{119}). 

Using momentum conservation we eliminate $p_2$, $p_2 = - \t p_1 + Nq - p_3$, so that, 
\bea \nn
\t p_1 {\cdot} p_2 &=& - 2+ N \t p_1 {\cdot} q - \t p_1 {\cdot} p_3\\ \nn
p_3 {\cdot} p_2 &=& -2 - \t p_1 {\cdot} p_3 + N p_3 {\cdot } q \\  q{\cdot} p_2 &=& - \t p_1{ \cdot }q - p_3 {\cdot} q~,
\eea
where we used that for tachyons $p^2= -m^2 =2$. 
We also eliminate $\t p_1 \cdot p_3$ by using $p_2^2 = (\t p_1 - N q + p_3)^2$, which gives, 
\be
\t p_1 {\cdot }p_3 = N \t p_1 {\cdot} q + N p_3{ \cdot} q -1~.
\ee
As a result, 
\bea \nn
\t p_1 {\cdot} p_2  &=& - 1 - N p_3 {\cdot} q~ \nn \\ p_3 {\cdot} p_2 &=& -1 - N \t p_1 {\cdot} q \nn \\ q\cdot p_2 &=& - \t p_1{ \cdot }q - p_3{ \cdot }q~.
\eea
Inserting into the amplitude (\ref{118}) gives
\bml
\!\!\!\!\!\!\!\mA = \frac{1}{\text{vol}(SL_2)}\!\!\int \frac{d w_1 d w_2 d w_3}{w_{12} w_{13} w_{23}}\!\prod_a  d z_a\!\(\!\frac{\!w_{13} (w_{2}-z_a)}{w_{12} (w_3-z_{a})}\!\)^{m_a p_3 \cdot q} \\
\(\frac{\!w_{13} (w_{2}{-}z_a)}{w_{23} (w_1{-}z_a)}\)^{m_a \t p_1 \cdot q}  \( - \sum_{i=1}^3 \frac{p_i {\cdot} \lam_a}{w_i-z_{a}}\)~.
\end{multline}
Using $p_1 \cdot \lam_a = 0$ (which is the true for  kinematics described in the main body of the text), the  last term simplifies to, 
\be
- \sum_{i=1}^3 \frac{p_i {\cdot} \lam_a}{w_i-z_{ a}} = -p_3 {\cdot} \lam_a\frac{w_{23}}{(w_3{-}z_{a})(w_2{-} z_{a})}~,
\ee
and the amplitude takes the form, 
\bml
\mA = \frac{-1}{\text{vol}(SL_2)}\int \frac{d w_1 d w_2 d w_3 }{w_{12} w_{13} }\, \prod_a \frac{d z_a  (p_3\cdot \lam_a )}{(w_2-z_{a})(w_3 -z_{a})} \\ \(\frac{w_{13} (w_2{-}z_{ a})}{w_{12}(w_3{-}z_{a})}\)^{m_a p_3 \cdot q} \!
 \(\frac{w_{13} (w_{2}{-}z_a)}{w_{23} (w_1{-}z_a)}\)^{m_a \t p_1 \cdot q}  \!\!
\end{multline}
We need to sum over all ordering of the points $w_i,z_a$. The only relevant orderings for us are those in which the $z_a$ can collide with $w_1$ (these are the ones that will have poles corresponding to  photons  scattering off of the initial tachyon). So we take the ordering $w_1<\{z_a\}<w_2<w_3$. We are indifferent to the orderings of the $z_a$ among themselves, and we will sum over all of them. Using SL$_2$ symmetry, we set $w_1 = 0$, $w_2 = 1$, $w_3 = \infty$. Because we are summing over all orderings of the $z_a$ among themselves, the integrals over $z_a$ factorize for different $a$. We get that $\mA$ (up to an irrelevant sign) is, 
\bea \nn
&&\prod_a (p_3\cdot \lam_a)\! \!\int_0^1\! d z_a ( 1{-}z_a)^{m_a (p_3 \cdot q + \t p_1 \cdot q) - 1} z_a^{- m_a \t p_1 \cdot q}\\ \nn
 &&= \prod_a( p_3\cdot \lam_a)  \frac{\G(m_a(p_3 \!\cdot\! q {+} \t p_1 \!\cdot\! q)) \G(- m_a\t p_1\! \cdot\! q {+} 1)}{\G(m_a p_3 \!\cdot\! q {+}1)}
\eea
Expanding near the pole at $\t p_1 \cdot q= 1$ we obtain (\ref{eq11}), up to a prefactor. 
\\

\section{Appendix B: The Baker's Map}

One of the simplest chaotic systems
is the baker's map, a two-dimensional discrete-time map. Phase space is a square, $0\leq x, y<1$. The map stretches phase space  by a factor of two in the $x$ direction, contracts it by a factor of two in the $y$ direction, and then the right $x$ half is stacked on top of the left half, to get a square again. Explicitly, the map is \cite{Dorfman},
\be \label{BS}
(x_{t+1},y_{t+1}) = \begin{cases}  (2 x_t, \frac{y_t}{2})  \ \ \ \ \ \ &0\leq x_t\leq \frac{1}{2} \\
 (2 x_t- 1, \frac{y_t + 1}{2} ) & \frac{1}{2}< x_t\leq 1~. \end{cases}
 \ee
 The map achieves the hallmarks of classical chaos -- stretching and folding of phase space -- in the most direct possible way. The initial conditions can be written in binary form,
\be
x_0 = \sum_{n=1}^{\infty} \frac{c_{-n}}{2^n}~, \ \ \ \ \ y_0 
= \sum_{n=1}^{\infty} \frac{c_{n-1}}{2^n}~,
\ee
with coefficients $c_n$ which are either zero or one. Application of the map $t$ times gives $(x_t, y_t)$, 
\be \label{xtB}
x_t = \sum_{n=1}^{\infty} \frac{c_{-n-t}}{2^n}~, \ \ \  \ \ y_t 
= \sum_{n=1}^{\infty} \frac{c_{n-1-t}}{2^n}~.
\ee
If we represent the location $(x,y)$ as $\cdots c_{-3} c_{-2} c_{-1}. c_{0} c_1 c_2 \cdots$, then after $t$ time steps, the decimal shifts to the left by $t$ units.

To see  chaos, consider a generic initial state: a randomly chosen $(x_0,y_0)$, picked with uniform measure  within the unit square (equivalently, each $c_n$ for each integer $n$ is picked to be either $0$ or $1$, with equal measure for both). The position $x_t$ will appear erratic as a function of time $t$. Alternatively, consider a change in $x_0$, by  for instance flipping the value of $c_{-n}$ for some particular  $n$; if $n$ is large and positive, the change in $x_0$ is small. From (\ref{xtB}) one sees that the change in $x_t$ becomes of order one for times $t$  of order $n$. \\

There is a heuristic (partial) analogy between the baker's map and the decay of an excited string into two light strings. The excited string is specified by a set of mode occupation number $\{n_m\}$, taking arbitrary values $n_m = 0,1,2\ldots $ for modes $m =1,2, 3, \ldots$.
The amplitude is a function of the angle $\al$ and is of the form (\ref{mAal}). As noted earlier, for a generic excited state of the string, $\{n_m\}$,  the amplitude appears erratic as a function of $\al$ and, in addition, a small change in the state leads to a large change in the amplitude. 

The time $t$, position $x_t$, and initial conditions $\{c_n\}$ in the baker's map are analogous to the angle $\al$, amplitude $\mA_{H\rightarrow T + T}(\a)$, and mode occupation numbers $\{n_m\}$, respectively, in the string decay amplitude. Both these maps input a set of integers and output a dynamical, chaotic quantity: the position $x_t$ (\ref{xtB}) for the baker's map, and the amplitude $\mA_{H\rightarrow T + T}(\al)$ (\ref{mAal}) for the string decay. 

The analogy only goes so far: while for the baker's map there is a straightforward relation between the change in a coefficient $c_{-n}$ and the amplitude at a particular time, for the string decay amplitude - which is a product rather than a sum - exciting some mode $m$ causes a significant change in the amplitude at multiple angles $\al$.

\bibliographystyle{utphys}

\end{document}